\pdfoutput=1

\documentclass[twocolumn]{bmcart}

\usepackage[utf8]{inputenc} 

\usepackage{graphicx}
\usepackage{amsmath}
\usepackage{amssymb}

\startlocaldefs
\newcommand{\dtbbx}{x}
\newcommand{\ctbbx}{\tilde{x}}
\newcommand{\ctpbx}{\hat{x}}
\newcommand{\ctpbxamp}{x}
\newcommand{\dtbbrx}{r}
\newcommand{\ctbbrx}{\tilde{r}}
\newcommand{\ctpbrx}{\hat{r}}

\newcommand{\dacnl}[2]{\alpha _{#1,#2}}
\newcommand{\ampnl}[1]{\beta _{#1}}
\newcommand{\txmixergain}{\gamma _{\text{Tx}}}
\newcommand{\txmixergainconj}{\delta_{\text{Tx}}}
\newcommand{\rxmixergain}{\gamma _{\text{Rx}}}
\newcommand{\rxmixergainconj}{\delta_{\text{Rx}}}
\newcommand{\eqmixergain}{\gamma _{\text{Eq}}}
\newcommand{\eqmixergainconj}{\delta_{\text{Eq}}}
\endlocaldefs

\begin{document}

\begin{frontmatter}

\begin{fmbox}
\dochead{Research}





\title{Baseband and RF Hardware Impairments in Full-Duplex Wireless Systems: Experimental Characterisation and Suppression}



\author[
   addressref={aff1},                   
   corref={aff1},                       
   email={alexios.balatsoukas@epfl.ch}   
]{\inits{ABS}\fnm{Alexios} \snm{Balatsoukas-Stimming}}
\author[
   addressref={aff1},
   email={andrew.austin@epfl.ch}
]{\inits{ACMA}\fnm{Andrew C. M.} \snm{Austin}}
\author[
   addressref={aff1},
   email={pbelanov@gmail.com}
]{\inits{PB}\fnm{Pavle} \snm{Belanovic}}
\author[
   addressref={aff1},
   email={andreas.burg@epfl.ch}
]{\inits{AB}\fnm{Andreas} \snm{Burg}}


\address[id=aff1]{
  \orgname{Telecommunications Circuits Laboratory, EPFL}, 
  \street{STI-IEL-TCL, Station 11},                     %
  \postcode{CH-1015}                                
  \city{Lausanne},                              
  \cny{Switzerland}                                    
}





\begin{abstractbox}

\begin{abstract} 
Hardware imperfections can significantly reduce the performance of
full-duplex wireless systems by introducing non-idealities and random
effects that make it challenging to fully suppress self-interference.
Previous research has mostly focused on analyzing the impact of hardware imperfections on full-duplex systems, based on simulations and theoretical models. In this
paper, we follow a measurement-based approach to experimentally
identify and isolate these hardware imperfections leading to residual
self-interference in full-duplex nodes. Our measurements show the
important role of images arising from in-phase and
quadrature (IQ) imbalance in the mixers. We also observe
base-band non-linearities in the digital-to-analog converters (DAC), which can
introduce strong harmonic components that have not been previously
considered. A corresponding general mathematical model to suppress these components of the
self-interference signal arising from the hardware non-idealities is
developed from the observations and measurements. Results from a
10~MHz bandwidth full-duplex OFDM system, operating at 2.48~GHz, show
up to 13~dB additional suppression, relative to state-of-the-art
implementations can be achieved by jointly compensating for IQ
imbalance and DAC non-linearities.


\end{abstract}




\end{abstractbox}
\end{fmbox}

\end{frontmatter}




\section{Introduction}
The increasing demand for wireless communications has renewed interest in developing systems that more efficiently use the limited radio spectrum to increase throughput, support additional users, and improve capacity. A promising physical-layer approach to increase the spectral utilisation is full-duplex operation, where transceivers in the system transmit and receive simultaneously in the same frequency band~\cite{Chen1998,Sabharwal2014,Hong2014}. It is important to note that most contemporary wireless communication systems use half-duplex, separating transmission and reception in frequency or time. Accordingly, full-duplex systems can potentially double the spectral efficiency~\cite{Sabharwal2014,Jain2011,Duarte2012} and are of considerable interest for cognitive radio~\cite{Afifi2013,Zheng2013}, co-operative communications~\cite{Krikidis2012,Zhang2014}, relay networks~\cite{Hiep2014} and 5G wireless systems~\cite{Hong2014}.  Furthermore, full-duplex links can be used to increase physical layer secrecy~\cite{Zheng2013a}, improve medium access control (MAC) layer protocols to increase throughput~\cite{Oashi2012}, and simplify resource allocation and spectrum management~\cite{Sabharwal2014}. 

One of the main challenges for realizing full-duplex systems is the presence of strong self-interference arising from physically close transmitting and receiving antennas~\cite{Chen1998,Jain2011,Duarte2012}. A full-duplex system can also be implemented using only a single antenna and a circulator, however, imperfect isolation will still allow self-interference signals to leak between the transmit and receive circuits. The power of the self-interference signal can be orders of magnitude larger than any external signal-of-interest, and full-duplex systems must therefore implement self-interference suppression~\cite{Sabharwal2014}. Ideally, this suppression will reduce the power of the self-interference signal to or below the noise floor. Where this cannot be achieved, the residual self-interference will lower the signal-to-noise ratio (SNR) and reduce throughput.

In principle, perfectly suppressing the self-interference signal should be possible, as the baseband transmitted signal is always known within the full-duplex node. However, previous research has shown the actual self-interference signal is a complicated function of the baseband transmitted signal, which is altered through many effects that are only partially understood or known~\cite{Sabharwal2014}.  
The impact and mitigation of such effects have been considered extensively in the literature for conventional wireless communication systems, for example~\cite{Fettweis2005, Schenkbook2008, SmainiBook2012}. However, full-duplex systems require a more detailed characterisation and modelling of these non-idealities due to the considerable power of the self-interference signal compared to the desired signal. Hence, hardware imperfections are still a significant, if not the dominant limiting factor for analog and digital self-interference suppression techniques~\cite{Sabharwal2014,Zheng2013}. Previous research into the effects of hardware imperfections for full-duplex systems have examined non-linearities introduced by the analog-to-digital converters (ADC)~\cite{Korpi2014b} and RF power amplifiers~\cite{Bharadia2013,Korpi2014,Li2014}, oscillator phase-noise~\cite{Shao2014,Sahai2013,Ahmedtobepublished,Syrjala2014}, in-phase and quadrature (IQ) imbalance~\cite{Korpi2014b,Li2014}, and sampling jitter~\cite{Syrjala2014a}. However, most of these studies are based on theoretical models and simulation analysis of assumed hardware imperfections without experimental verification. Unfortunately, it is mostly the uncertainty and inaccuracy of the models that renders the suppression of these effects difficult. Hence, their identification through measurements and experiments takes an important role in enabling full-duplex communication nodes.

\noindent\textit{Contributions and Outline:} The motivation for this paper is, therefore, to consider an experimental characterisation of the hardware imperfections present on a full-duplex platform to complement the existing body of theoretical studies describing their impact. Specifically, we confirm the existence and the importance of most transmitter impairments that have been previously studied theoretically in the literature. More importantly, we identify a strong baseband non-linearity coming from the digital-to-analog converters (DACs) which, to the best of our knowledge, has not received any attention in the existing full-duplex literature. While the measured values for the impairments are specific to our hardware platform, the findings are generalised to a parametric cancellation model that can be applied to any similar full-duplex system. This digital cancellation model jointly takes into account DAC non-linearities and IQ imbalance and outperforms existing digital cancellation methods. 

The paper is organised as follows. Section~\ref{sec:fullduplex} outlines the various self-interference suppression architectures and their susceptibility to hardware imperfections. A brief description of our hardware testbed is provided in Section~\ref{sec:testbed}. In Section~\ref{sec:impairments} we measure and model the non-ideal effects introduced by the hardware using single tone tests. Section~\ref{sec:cancelling} shows the impact of correctly compensating for the hardware effects identified in the previous sections for a 512-tone \textsc{OFDM}-like full-duplex transceiver.

\section{Self-Interference Suppression in Full-Duplex Wireless Systems} \label{sec:fullduplex}

Physical separation and inherent attenuation between the transmit and receive circuits introduces a small amount of passive suppression. However, typically passive suppression (or the isolation through a circulator) alone is insufficient to allow reliable full-duplex communications~\cite{Everett2014}. Additional signal processing is required to actively suppress and reduce the strength of the self-interference to the noise-floor. In general, some of this active suppression must be achieved \textit{before} the received signal is digitised, i.e., in the analog domain, since entirely digital suppression is usually not feasible due to limited ADC resolution when the external signal-of-interest is small relative to the self-interference~\cite{Sabharwal2014}. 

\begin{figure}[t]
	\centering
	\includegraphics[width=0.42\textwidth]{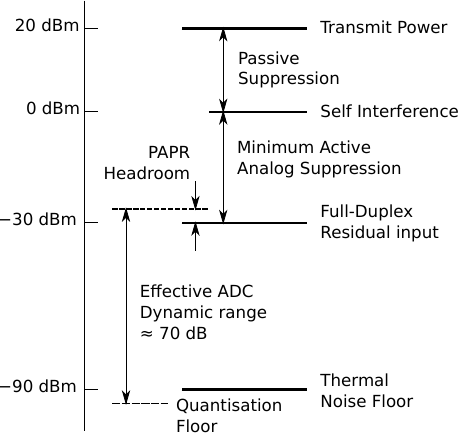}
	\caption{Power budget for a full-duplex wireless system showing the various stages required to suppress the transmitted signal. }
	\label{fig.power.levels}
\end{figure}

For example, Fig.~\ref{fig.power.levels} shows the required suppression budget for a 10~MHz full-duplex wireless system, with a maximum transmit power of 20~dBm. For this operating bandwidth the typical measured noise floor is approximately $-90$~dBm. The peak to average power ratio (PAPR) of the 512-tone OFDM signal is approximately 10~dB, and we must therefore allow a similar amount of head-room to avoid overloading the receiver front-end circuitry. To prevent the system from being limited by the ADC quantisation, it is necessary to place the quantisation-floor at or below the thermal-noise floor. At 20~dBm transmit power, a minimum of 50~dB suppression is required from the combined passive and analog suppression stages to reduce the power of the self-interference to a level where the remaining signal can be captured within the dynamic range of the ADC, with a sufficient resolution for the desired signal after digital-domain suppression of the residual self-interference. 

\subsection{Active Analog Suppression}
One of the most widely considered and successful active analog suppression techniques is to subtract a cancellation signal from the received signal (which contains the self-interference and any desired signal)~\cite{Chen1998,Jain2011,Duarte2012}. A number of techniques for generating this cancellation signal have been proposed, however, these generally fall into two categories: 

The \textit{Stanford} architecture~\cite{Jain2011,Bharadia2014} proposes measuring the transmitted RF signal (e.g., with a low insertion-loss coupler) immediately before it is applied to the antenna. The cancellation signal is generated by appropriately delaying and attenuating this measured signal to account for propagation in the real self-interference channel. A practical implementation of this approach using tapped delay lines achieved $45$~dB active suppression, over a $40$~MHz bandwidth at $2.45$~GHz~\cite{Jain2011}. An advantage of the Stanford architecture is that hardware imperfections introduced by the transmitter circuit are inherently included in the cancellation signal. However, system-specific RF cancellation circuits are required to model the effect of the channel and the architecture does not scale well to Multiple-Input Multiple-Output (MIMO) systems, as each transmitter-receiver pair requires a dedicated circuit~\cite{Bharadia2014}.  

In contrast, the analog cancellation signal in the \textit{Rice} architecture is generated using a separate RF chain~\cite{Duarte2010,Duarte2012}. The baseband input to the cancellation chain is computed by sounding the self-interference channel to determine the appropriate delay, attenuation, and distortion parameters that must be applied. Several implementations of the Rice architecture have been reported~\cite{Duarte2012,Balatsoukas-Stimming2013}. The initial prototype achieved between 20--34~dB analog suppression for transmit powers ranging from 0--15~dBm (generally suppression decreases with increasing transmit power), over a 40~MHz bandwidth on a WARP platform~\cite{Duarte2012}. Similar results were reported using a National Instruments~(NI) FlexRIO platform, with an average 48~dB active suppression over a 20~MHz bandwidth for 4~dBm transmit power~\cite{Balatsoukas-Stimming2013}. For both implementations, further suppression was achieved in the digital domain, however, both groups observed the suppression introduced by digital cancellation depended on the analog stage, and that the total suppression achieved was approximately constant~\cite{Duarte2012,Balatsoukas-Stimming2013,Sahai2013}.  The Rice architecture can be more readily extended to MIMO systems (as each receiver only requires one additional RF cancellation chain) and does not require specialised RF hardware design. However, it is important to note that hardware imperfections are potentially more prevalent in the Rice architecture, as the cancellation signal is generated from a separate RF chain.

\subsection{Active Digital Suppression}
The active analog suppression stage, which is usually not perfect, can be followed by suppression in the digital domain to further remove remaining self-interference. In general, i.e., with or without active analog suppression, the received complex baseband signal, $r$, can be expressed as a sum of: the external signal-of-interest, $s$; a function, $f(\cdot)$ of the complex baseband transmitted self-interference signal, $x$; and noise, $z$. The goal of digital suppression is to estimate $f(\cdot)$ and subtract $f(x)$ from the received signal.

Since the transmitted signal is distorted by transmitter non-idealities, the self-interference signal is a complicated function of $\dtbbx$. Sophisticated digital cancellation methods are required to capture those transmitter imperfections~\cite{Bharadia2013,Korpi2014}. A thorough identification and characterisation of these impairments is presented in Section~\ref{sec:impairments}, and digital cancellation schemes are presented in Section~\ref{sec:cancelling}.

\section{System Architecture and Testbed Setup} \label{sec:testbed}

\begin{figure}[t]
	\centering
	\includegraphics[width=0.44\textwidth]{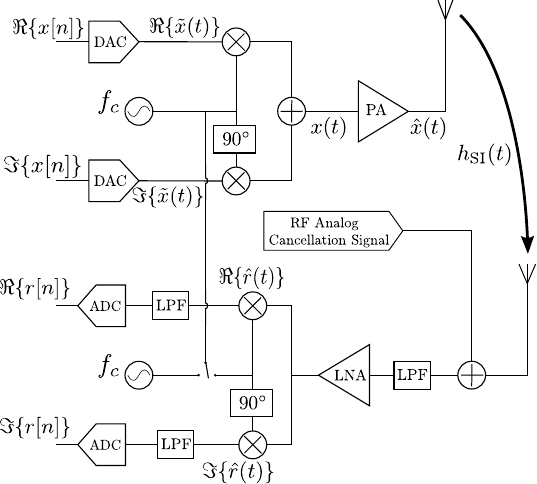}
	\caption{Block diagram of a full-duplex transceiver node with all important analog front-end components.}\label{fig:bd}
\end{figure}

\begin{figure}[t]
	\centering
	\includegraphics[width=0.44\textwidth]{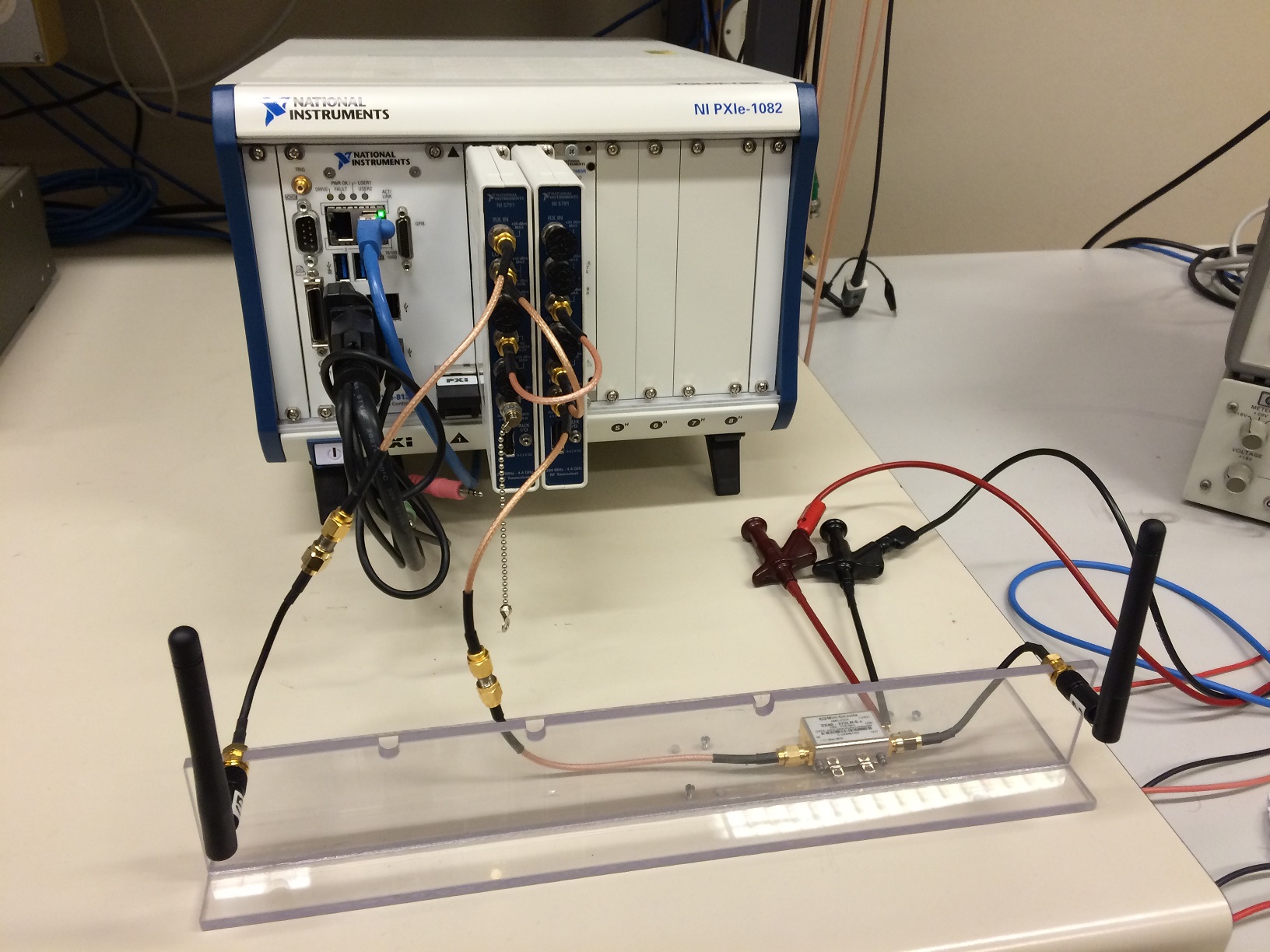}
	\caption{Full-duplex front-end showing the antenna configuration considered and the NI FlexRIO hardware platform with two NI~5791R RF transceiver adapter modules installed.}\label{fig:photo}
\end{figure}

Our full-duplex node testbed (running the Labview software platform) consists of a NI FlexRIO PXIe-1082 chassis~\cite{NIPXIe-1082} with two NI~5791R RF transceiver adapter modules~\cite{NI5791R}, each containing one transmitter and one receiver. Each NI~5791R module inherently uses the same oscillator for the receiver and the transmitter, and the carrier signal is also shared between multiple cards. The NI~5791R operates at carrier frequencies of $400$~MHz up to $4.4$~GHz and provides output powers ranging from $-24$~dBm to $8$~dBm. In order to be able to test realistic output powers of up to $22$~dBm, we use an external amplifier~\cite{MCZXZX60-272LN+} that provides $14$~dB of gain. A photograph of the system is shown in Fig.~\ref{fig:photo}. 

The digital baseband transmit samples are generated in MATLAB and sent to the testbed over a network. The testbed is responsible for synchronizing the receiver and transmitter and handles all analog frontend tasks (i.e., digital-to-analog conversion, mixing, amplification, RF filtering, and analog-to-digital conversion). The received digital baseband samples are then sent back to MATLAB for offline processing. For the measurements in this paper we use an RF frontend that consists of two $2.4$~GHz antennas with variable spacing and orientation.

In this paper, we use this two-antenna configuration to emulate the analog suppression by increasing the level of passive isolation. 
While our analog suppression achieves the necessary 50\,dB attenuation even with a circulator, we are interested in observing \emph{all} 
signal components in the digital domain for our measurements and experiments. Hence, in the following, we use the system with the active analog suppression 
chain deactivated. 

\section{Transceiver Impairments} \label{sec:impairments}
In this section, we discuss the main sources of non-idealities in the self-interference signal and we provide measurements from our testbed that clearly demonstrate the existence of most of the impairments that have been previously considered in the bibliography. More specifically, we confirm the presence and the effect of phase noise~\cite{Shao2014,Sahai2013,Syrjala2014,Ahmedtobepublished}, IQ imbalance~\cite{Korpi2014b,Li2014}, and RF non-linearities~\cite{Bharadia2013,Korpi2014,Li2014}. More importantly, however, we observe the existence of baseband non-linearities with significant power. We show that these  non-linearities are highly consistent with non-linearities stemming for the transmitter DACs, and we provide a corresponding DAC non-linearity model that can be used for digital self-interference cancellation.

We denote the discrete time digital baseband signal by $\dtbbx[n]$ and the continuous-time analog baseband signal by $\ctbbx(t)$. The upconverted analog signal is denoted by $\ctpbx(t)$ and the amplified upconverted analog signal is denoted by $\ctpbxamp(t)$. The downconverted analog self-interference signal at the receiver is denoted by $\ctbbrx(t)$ and the digital baseband self-interference signal is denoted by $\dtbbrx[n]$. For simplicity, we ommit the time indices $n$ and $t$, unless strictly necessary (e.g., to denote a delay), as they will always be clear from the context.

\subsection{Phase Noise}\label{sec:phasenoise}
The upconversion of the baseband signal to the carrier frequency $f_c$ is performed at the transmitter by mixing the baseband signal with a carrier signal. The oscillators that are used to generate the carrier signal suffer from various impairments, the most significant of which is phase noise. Thus, instead of generating a pure tone at frequency $f_c$, i.e., $e^{j2\pi f_c t}$, the generated tone is actually $e^{j(2\pi f_c t+\phi(t))}$, where $\phi(t)$ is the random phase noise process. The downcoversion process at the receiver is also affected by phase noise, since it uses a similarly generated carrier signal. 

The effect of phase noise on full-duplex transceivers has been extensively studied~\cite{Sahai2013,Syrjala2014,Ahmedtobepublished}. In order to summarize and illustrate the effect of the phase noise, we assume for the moment that all parts of the transceiver are ideal, except for the oscillators that generate the carrier signal. Moreover, for illustration purposes, the self-interference channel is assumed to introduce a simple delay, i.e., it can be represented as $\delta (t-\Delta t)$. Such a delay can for example arise from accoustic wave bandpass filters in the receive chain. At the transmitter, phase noise is introduced during the upconversion process, so the transmitted RF signal is:
\begin{align}
	\ctpbx	& = \Re\left\{\ctbbx e^{j(2\pi f_ct+\phi_{\mathrm{Tx}}(t))}\right\},
\end{align}
where $\phi_{\mathrm{Tx}}(t)$ denotes the phase noise process of the oscillator used by the transmitter. At the receiver, phase noise is introduced during the downconversion process:
\begin{align}
	\ctpbrx	& = \text{LPF}\left\{\ctpbx e^{-j(2\pi f_ct+\phi _{\mathrm{Rx}}(t-\Delta t))}\right\} \\
					& = \ctbbx e^{j(\phi_{\mathrm{Tx}}(t)-\phi _{\mathrm{Rx}}(t-\Delta t))-2\pi f_c \Delta t},
\end{align}
where $\text{LPF}$ denotes a low-pass filter that removes the copies of the signal around $2f_c$ and $-2f_c$ and $\phi_{\mathrm{Rx}}(t)$ denotes the phase noise process of the oscillator used by the receiver. If the transmitter and the receiver use independent oscillators, the $\phi_{\mathrm{Tx}}(t)$ and $\phi_{\mathrm{Rx}}(t)$ processes will be uncorrelated. However, in full-duplex transceivers the transmitter and the receiver are typically co-located and can physically share the same oscillator. Thus, $\phi_{\mathrm{Tx}}(t) = \phi_{\mathrm{Rx}}(t)$ and we denote the common phase noise process by  $\phi (t)$. Due to the delay introduced by the transmission channel, the phase noise instances experienced by the signal at the transmitter and the receiver mixer are not identical. However, it is evident that if the delay is such that $\phi(t)$ and $\phi (t-\Delta t)$ are highly correlated, then sharing the oscillator can significantly reduce the effect of phase noise in the received signal after the mixer~\cite{Sahai2013}.

\begin{figure}[t]
	\centering
	\includegraphics[width=0.44\textwidth]{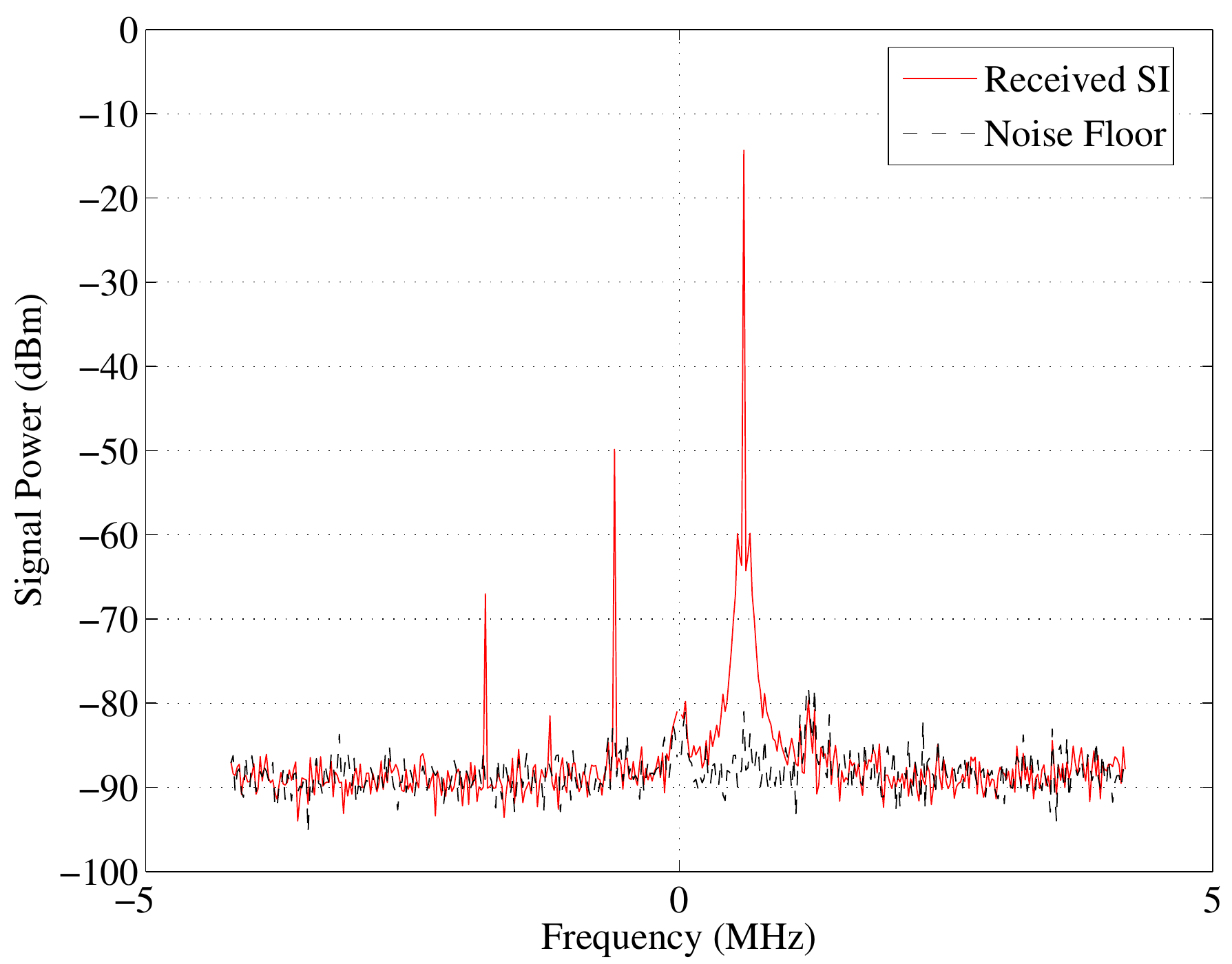}
	\caption{Received self-interference signal ($-10$~dBm).}\label{fig:phasenoise}
\end{figure}

\begin{figure}[t]
	\centering
	\includegraphics[width=0.44\textwidth]{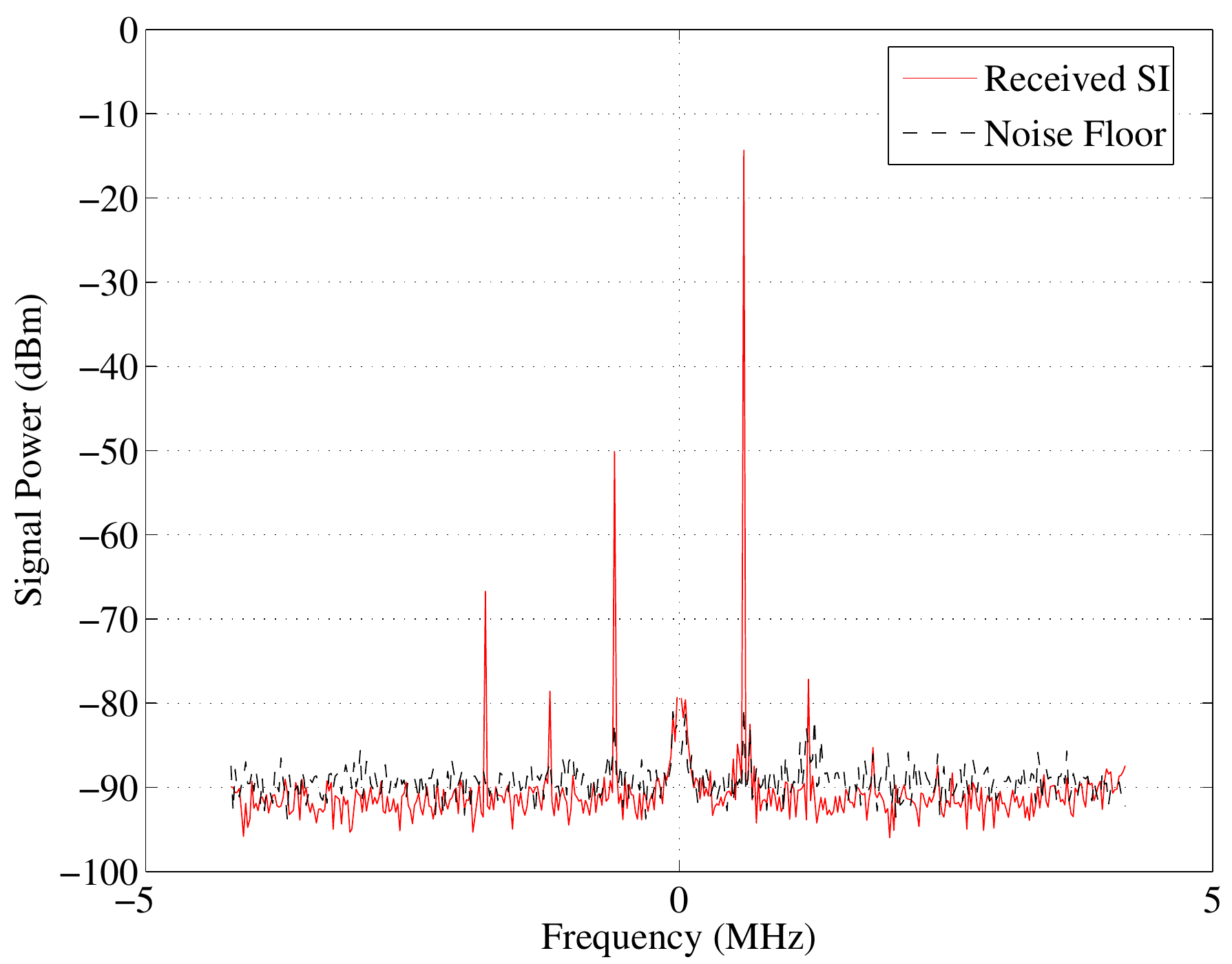}
	\caption{Received self-interference signal with shared oscillator ($-10$~dBm).}\label{fig:phasenoise2}
\end{figure}

In order to demonstrate the improvement obtained by sharing the oscillator between the transmitter and the receiver mixers, we perform a one-tone test on our testbed and we examine the received self-interference signal. We use the lowest possible transmit power setting (i.e., $-10$~dBm) in order to minimize non-linearities arising from the amplifier of the transmitter. An indicative spectrum of the received self-interference signal is presented in Fig.~\ref{fig:phasenoise}. We observe that the received signal has significant spectral content around the transmitted tone, with the most powerful components lying approximately $46$~dB below the power of the tone. We note that the two independent oscillators use the same $10$~MHz reference signal, but this is not sufficient to reduce the effect of phase noise since it is caused by the phase-locked loop (PLL) that generates the actual carrier signal from the reference signal. We also observe numerous tones arising from other non-linearities, which will be explained in the following sections. In Fig.~\ref{fig:phasenoise2}, we present an indicative spectrum of the received self-interference signal when the transmit and receiver mixers use the same oscillator. In this case, the strongest component of the spectral content resulting from phase noise lies approximately $70$~dB below the received tone. We observe that the phase noise induced noise floor will lie significantly below the noise floor introduced by the remaining non-linearities.

\subsection{Baseband Non-Linearities}\label{sec:bbnl}
In Fig.~\ref{fig:phasenoise}, we observe that numerous tones have appeared in the received signal apart from the transmitted tone. The eminent tones on the positive side of the spectrum appear at integer multiples of the transmitted tone frequency. Since the transmitter amplifier is set to its lowest possible setting and, more importantly, we observe \emph{even} harmonics of the transmitted tone, we can safely conclude that these harmonics must (at least partially) occur in the analog baseband signal. The only components that can introduce non-linearities in the transmitter-side analog baseband signal are the two DACs. On the receiver side, we have two ADCs, which can also introduce non-linearities in the observed digital baseband signal. However, the ADC used in the NI 5791R transceiver has a higher spurious-free dynamic range (SFDR) than the DAC, so we assume that all baseband non-linearities stem from the DACs.

We model the DAC-induced non-linearities by using a Taylor series expansion around $0$ of maximum degree $m_{\max}$. In the block diagram of Fig.~\ref{fig:bd}, we see that the first DAC has $\Re\{\dtbbx\}$ as its input and the second DAC has $\Im\{\dtbbx\}$ as its input. Thus the output signal of each DAC can be written as:
\begin{align}
	\Re\{\ctbbx\}	& = \sum _{m=1}^{m_{\max}} \dacnl{1}{m} \Re\{\dtbbx\} ^{m}, \\
	\Im\{\ctbbx\}	& = \sum _{m=1}^{m_{\max}} \dacnl{2}{m} \Im\{\dtbbx\} ^{m},
\end{align}
where $\dacnl{i}{m} \in \mathbb{R},~i\in \{1,2\},~m\in\{1,\hdots,m_{\max}\}$. Thus, the continuous time complex baseband signal $\ctbbx$ can be written as:
\begin{align}
	\ctbbx = \sum _{m=1}^{m_{\max}} \dacnl{1}{m} \Re\{\dtbbx\} ^{m} + j \sum _{m=1}^{m_{\max}} \dacnl{2}{m} \Im\{\dtbbx\} ^{m}. \label{eqn:dacnnl}
\end{align}
By analyzing \eqref{eqn:dacnnl} with a single input tone of frequency $f$, it can be shown that, if the DACs are perfectly matched so that $\dacnl{1}{m} = \dacnl{2}{m}~m\in\{1,\hdots,m_{\max}\},$ the DAC induced non-linearities produce harmonics alternatingly on only one side of the spectrum for odd $m$, but on both sides of spectrum for even $m$. More specifically, it is shown in the appendix that for odd $m$ we obtain harmonics at frequencies $m(-1)^{\frac{m-1}{2}}f$, while for even $m$ we obtain harmonics at both $-mf$ and $mf$ with equal power. We observe in Fig.~\ref{fig:phasenoise2} that the frequency $3f$ is not present but the frequency $-3f$ is present, and also that the harmonics at $-2f$ and $2f$ have approximately equal power. Thus, all our observations are in complete agreement with what we expect to see based on our model. The tone at frequency $-f$, which we observe clearly in Fig.~\ref{fig:phasenoise2} but is not predicted by the DAC non-linearities, is the result of IQ imbalance, as we will explain in the following section.

It is interesting to note that the tone at $-3f$ is stronger than the tone at $-2f$ in Fig.~\ref{fig:phasenoise2}, which seems counter-intuitive at first. However, when downscaling the digital baseband signal, we observe that the power of the 3rd harmonic decreases at a higher rate than the power of the 2nd harmonic, which is consistent with what one would expect.

\begin{figure}[t]
	\centering
	\includegraphics[width=0.44\textwidth]{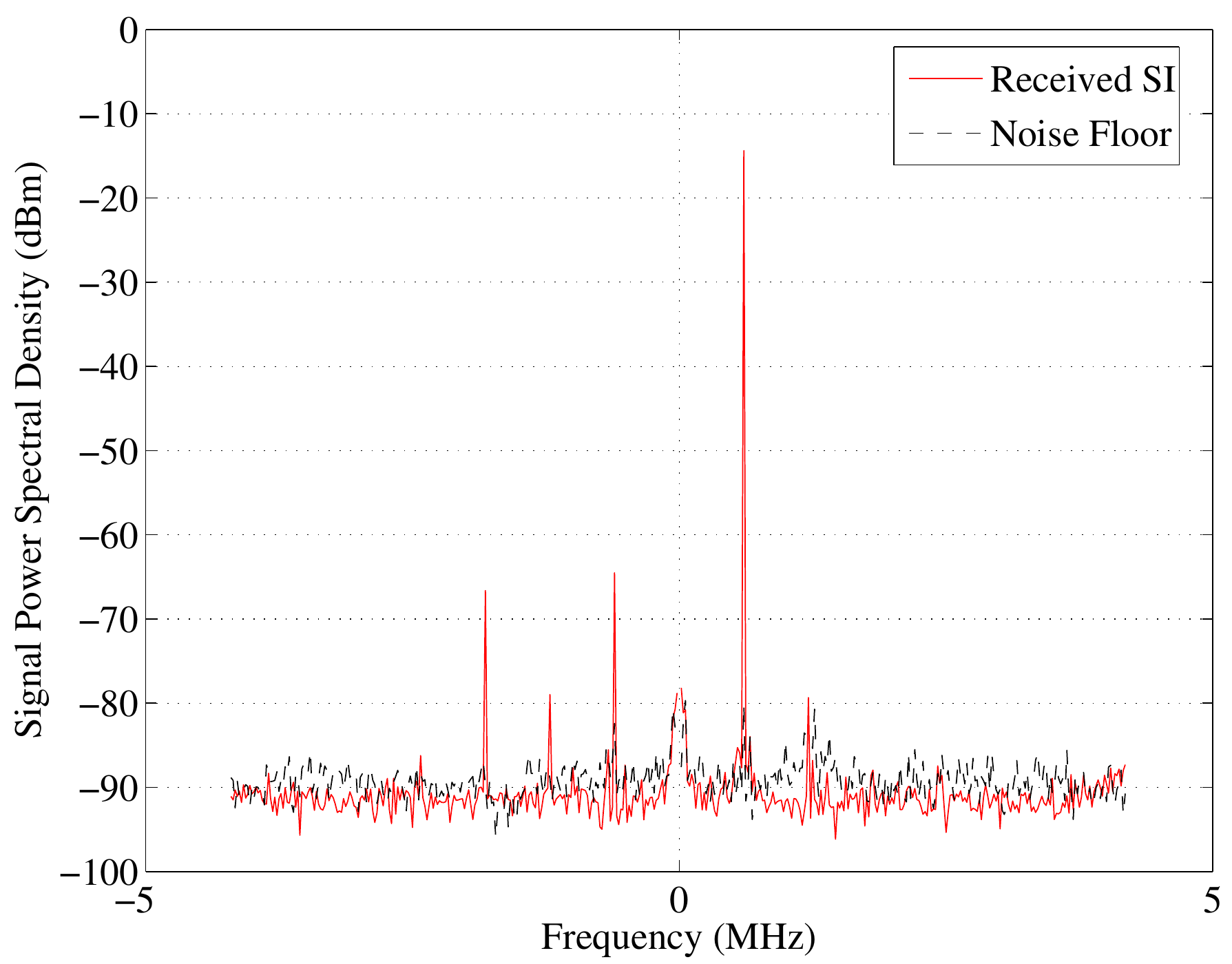}
	\caption{Received self-interference signal with shared oscillator and IQ imbalance compensation ($-10$~dBm).}\label{fig:IQcomp}
\end{figure}

\subsection{IQ Imbalance}
IQ imbalance is caused by amplitude and phase mismatch in the in-phase and the quadrature components of the upconverted analog signal. To simplify notation, in this section we consider frequency-flat IQ imbalance. The output of the non-ideal mixer can be modeled as:
\begin{align}
	\ctpbx	& = \Re\left\{\left(\txmixergain \ctbbx + \txmixergainconj \ctbbx ^*\right) e^{j2\pi f_ct}\right\},
\end{align}
where $\txmixergain, \txmixergainconj \in \mathbb{C}$. We note that any amplitude mismatch in the linear components of the DACs will also manifest itself as IQ imbalance.

In Fig.~\ref{fig:phasenoise}, we observe that there exists a mirror image of the transmitted tone with respect to the carrier frequency (i.e., at frequency $-f$), which arises due to the effect of IQ imbalance. 
However, it is important to note that the signal components that -at first sight- appear to be harmonics of this negative tone instead can only arise due to the DAC non-linearities as explained earlier. This is for several reasons: 
First, the harmonic of the original tone $f$ at frequency $3f$ is significantly weaker than the alleged harmonic of $-f$ at frequency $-3f$. Moreover, since there are no significant baseband non-linearities after the mixer of the receiver, we do not expect to observe a 2nd harmonic of $-f$ at frequency $-2f$. Finally, when we enable the built-in IQ imbalance compensation block of the NI 5791R transceivers, we observe that, while the power of the IQ imbalance induced tone at $-f$ is reduced by approximately $20$~dB, the apparent harmonics of this tone at $-2f$ and $-3f$ are unaffected, as demonstrated in Fig.~\ref{fig:IQcomp}. In order to have IQ imbalance that is similar to what a low-cost transceiver would experience, we keep the built-in IQ imbalance compensation mechanism of the NI~5791R disabled.

\begin{figure}[t]
	\centering
	\includegraphics[width=0.44\textwidth]{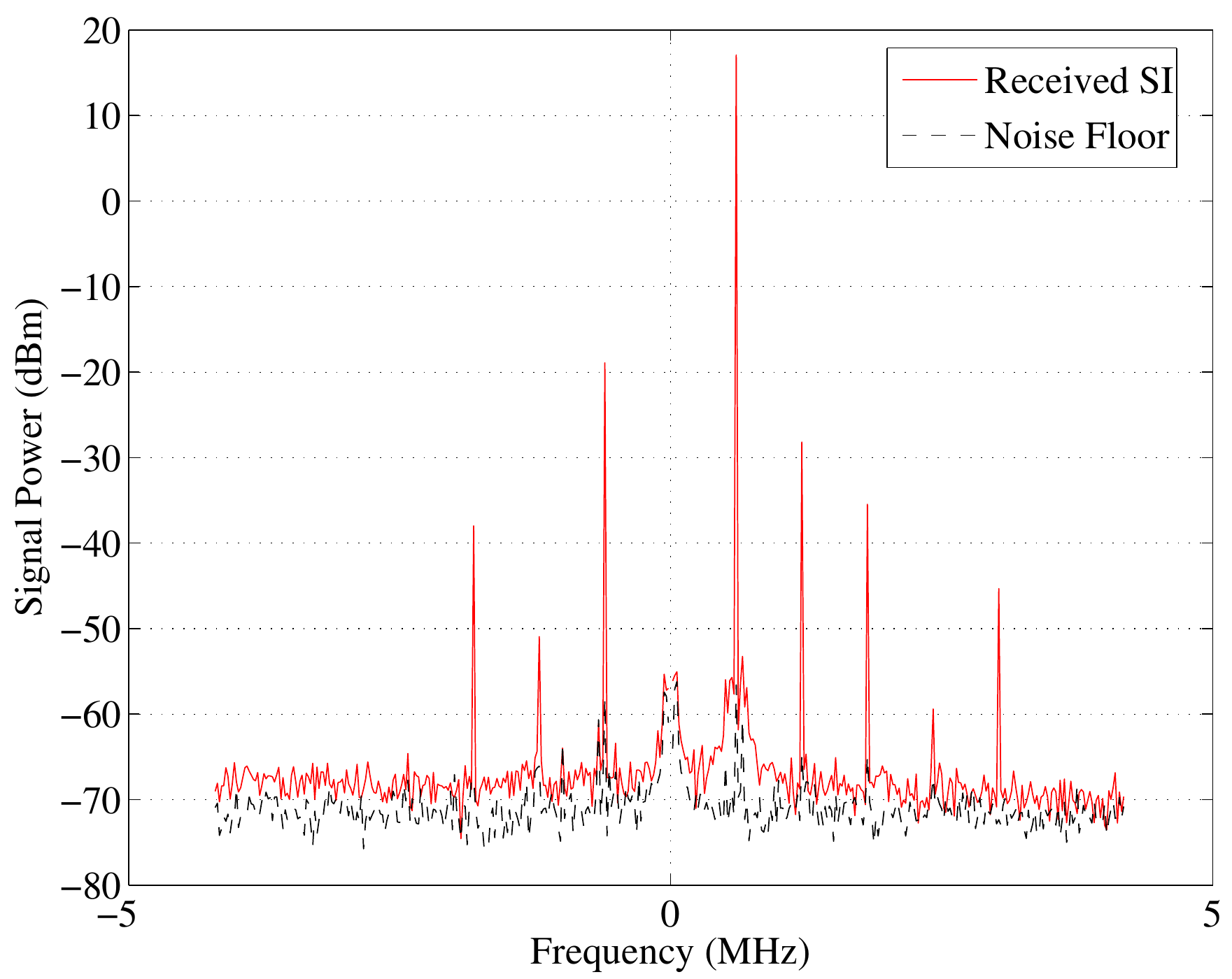}
	\caption{Received self-interference signal with shared oscillator ($20$~dBm).}\label{fig:RFnl}
\end{figure}

\subsection{RF Non-Linearities}
Non-linearities in the upconverted RF signal are caused by the power amplifier that comes after the RF mixer. These non-linearities mainly appear when the amplifier is operated in its non-linear region, i.e., close to its maximum output power, where significant compression of the output signal occurs. Basic arithmetic manipulations can show that all the even-power harmonics lie out of band and will be cut off by the RF low-pass filter of the receiver. The RF non-linearities can be modeled using a Taylor series expansion around $0$ of maximum degree $n_{\max}$:
\begin{align}
	\ctpbxamp &	= \sum _{\substack{n=1,\\n \text{ odd}}}^{n_{\max}}\ampnl{n}\ctpbx ^n, \label{eqn:ampnl}
\end{align}
where $\ampnl{n} \in \mathbb{R},~n\in\{1,3,\hdots,n_{\max}\}$.

The effect of RF non-linearities can be clearly seen in Fig.~\ref{fig:RFnl}, where we present the spectrum of the received self-interference signal when transmitting with an output power of $20$~dBm. We observe that strong 3rd and 5th harmonics of the transmitted tone $f$ appear. The tones on the negative frequencies remain almost unaffected, as expected because they do not arise from the RF non-linearities, but from the DAC non-linearities. The 3rd and 5th harmonics of the tones on the negative frequencies lie below the noise floor. We also observe that the noise floor has increased by $20$~dB. This is caused by the combined effect of the limited dynamic range of the ADCs of the receiver, which means that quantization noise dominates thermal noise. As can be seen by referring to the power budget in Fig.~\ref{fig.power.levels}, we would require at least $20$~dB more passive or active analog suppression in order to observe the thermal noise floor.

\section{Digitally Cancelling the Non-Idealities} \label{sec:cancelling}
As mentioned earlier, the goal of digital cancellation is to reconstruct the self-interference signal (including all transmitter non-idealities) and subtract it from the received signal. In this section, we briefly describe existing digital cancellation methods that take into account some of the transmitter impairments and we describe our proposed joint digital cancellation scheme.

\subsection{Existing Digital Cancellation Methods}
In general, i.e., with or without active analog suppression, the received complex baseband signal $r$ can be written as:
\begin{align}
	\dtbbrx & = f(\dtbbx) + s + z,
\end{align}
where $f(\dtbbx)$ denotes a function of the complex baseband self-interference signal $\dtbbx$, $s$ denotes the signal-of-interest and $z$ denotes thermal noise. The goal of digital suppression is to estimate the function $f$ and subtract $f(\dtbbx)$ from the received signal. In order to do so, $f$ needs to be modeled in some way.

The simplest form of digital cancellation, called \emph{linear cancellation}, models $f$ as a convolution with the self-interference channel, denoted by $h_{\textrm{SI}}$, i.e., 
\begin{align}
	\dtbbrx = h_{\textrm{SI}} \ast \dtbbx + s + z,
\end{align}
where $\ast$ denotes the convolution operation. By writing the convolution as a matrix operation, a least squares (LS) estimate for $h_{\textrm{SI}}$ can be obtained, which we denote by $\hat{h}_{\textrm{SI}}$. Linear digital cancellation can then be written as 
\begin{align}
	\dtbbrx - \hat{r}= h_{\textrm{SI}} \ast \dtbbx - \hat{h}_{\textrm{SI}} \ast \dtbbx + s + n,
\end{align}

In order to capture amplifier induced non-linearities, the \emph{non-linear} digital cancellation proposed in~\cite{Bharadia2013} assumes that
\begin{align}
	\dtbbrx = \sum _{\substack{n=1,\\n \text{ odd}}}^{n_{\max}}h_{\textrm{SI},n} \ast \dtbbx ^n + s + z,
\end{align}
where $h_{\textrm{SI},n},~n=1,3,\hdots,n_{\max},$ denotes the self-interference channel experienced by each of the harmonics of the baseband signal. In this case, $n_{\max}$ channels are estimated using the LS method, in order to remove the baseband signal and harmonics thereof from the received signal. 

\emph{Widely linear} digital cancellation, which takes into account IQ imbalance, was proposed in~\cite{Korpi2014}, where it is assumed that
\begin{align}
	\dtbbrx = h_{\textrm{SI}} \ast \dtbbx + h_{\textrm{SI},\text{IQ}} \ast \dtbbx ^* + s + z,
\end{align}
where $h_{\textrm{SI}}$ and $h_{\textrm{SI},\text{IQ}}$ denote the channels experienced by the baseband signal and the complex conjugate of the baseband signal, respectively. The goal becomes to jointly estimate $h_{\textrm{SI}}$ and $h_{\textrm{SI},\text{IQ}}$ using the LS method.

\subsection{Joint Digital Cancellation of DAC non-linearities and IQ Imbalance}
Ideally, we would like to perform digital cancellation based on a model that includes all non-idealities. However, due to multiple non-linearities, a full model is highly complicated. Thus, we first examine the case where the output power is low, so that it can be safely assumed that there are no RF non-linearities. In this case, the resulting non-idealities model leads to a convenient cancellation method.

\subsubsection{Low RF Output Power}
At low RF output power, the main sources of non-idealities are the DACs and the RF mixers, which introduce non-linearities and IQ imbalance, respectively. In the general case, IQ imbalance is frequency selective, so that $\txmixergain, \txmixergainconj \in \mathbb{C}^L$, where $L$ is the length of the impulse response. Thus, the analog RF signal $\ctpbx$ is given by 
\begin{align}
	\ctpbx	& = \Re\left\{\left(\txmixergain \ast \ctbbx + \txmixergainconj \ast \ctbbx ^*\right) e^{j2\pi f_ct}\right\}.
\end{align}
The transmitter amplifier operates in its linear regime so that the amplified analog RF signal, denoted by $\ctpbxamp$, is identical to $\ctpbx$. On the receiver side, the RF mixer introduces IQ imbalance during downconversion:
\begin{align}
\ctpbrx	& = \text{LPF}\left\{\left(\rxmixergain \ast h _{\mathrm{SI}} \ast \ctpbxamp + \rxmixergainconj \ast h _{\mathrm{SI}}^* \ast \ctpbxamp ^*\right) e^{-j2\pi f_ct}\right\} \\
				& = \frac{1}{2}\left[\eqmixergain \ast \ctbbx + \eqmixergainconj \ast \ctbbx ^*\right]  \label{eqn:nonidealitieslowpower:1} \\
				& = \frac{1}{2}\left[ (\eqmixergain+\eqmixergainconj) \ast \sum _{m=1}^{m_{\max}} \dacnl{1}{m} \Re\{\dtbbx\} ^{m} \right. \nonumber \\
				& +  \left. (\eqmixergain-\eqmixergainconj) \ast \sum _{m=1}^{m_{\max}} \dacnl{2}{m} \Im\{\dtbbx\} ^{m} \right],  \label{eqn:nonidealitieslowpower:2}
\end{align}
where:
\begin{align}
	\eqmixergain & \triangleq \rxmixergain \ast h _{\mathrm{SI}} \ast \txmixergain + \rxmixergainconj \ast h _{\mathrm{SI}} ^* \ast \txmixergainconj ^*, \\
	\eqmixergainconj & \triangleq \rxmixergainconj \ast h _{\mathrm{SI}}^* \ast \txmixergain^* + \rxmixergain \ast h _{\mathrm{SI}} \ast \txmixergainconj.
\end{align} 
Thus, from~\eqref{eqn:nonidealitieslowpower:1} we see that, for low transmit powers, the combined effect of the transmitter and receiver IQ imbalance and the transmission channel is equivalent to the effect of a single IQ imbalance with parameters $\eqmixergain$ and $\eqmixergainconj$. Moreover, by rewriting \eqref{eqn:nonidealitieslowpower:1} as \eqref{eqn:nonidealitieslowpower:2}, it becomes clear that a form of non-linear cancellation is required even at low output powers due to the non-linearities introduced by the DACs.

\begin{figure*}[t]
	\begin{align}
		\ctpbxamp &	= \sum _{\substack{n=1,\\n \text{ odd}}}^{n_{\max}}\ampnl{n}\ctpbx ^n = \sum _{\substack{n=1\\n \text{ odd}}}^{n_{\max}}\frac{\ampnl{n}}{2^n} \left[\sum _{k=0}^n {n \choose k}\left[ \left(\txmixergain \ast \ctbbx +	\txmixergainconj \ast \ctbbx ^*\right) e^{j2\pi f_ct}\right]^{n-k}  \left[\left(\txmixergain^* \ast \ctbbx^* + \txmixergainconj ^* \ast \ctbbx \right) e^{-j2\pi f_ct}  \right]^k\right] \label{eqn:txampnl}
	\end{align}
\end{figure*}

We propose to cancel the IQ imbalance and DAC non-linearities jointly, by constructing an LS based non-linear cancellation scheme based on the model of \eqref{eqn:nonidealitieslowpower:2}, which contains only $\Re\{\dtbbx\}$ and $\Im\{\dtbbx\}$ and powers thereof.

\subsubsection{Performance of Digital Cancellation}
In order to assess the performance of different digital cancellation mechanisms on our testbed, we conduct the following experiment. We construct $100$ OFDM-like frames containing $512$ tones spread over a $10$~MHz bandwidth and modulated with $4$-QAM symbols. We transmit each group of frames using transmit powers ranging from $-10$~dBm up to $22$~dBm. The carrier frequency is set to $2.48$~GHz. We consider two antenna spacings that give us $40$~dB and $55$~dB of passive analog suppression. In practice these suppression numbers are easily achievable by using a combination of passive suppression and active analog suppression, but as mentioned earlier, in our experiments no active analog cancellation is performed to keep all signal components accessible. The digital baseband samples are recorded and the various digital cancellation methods are applied to them off-line.

\begin{figure}[t]
	\centering
	\includegraphics[width=0.44\textwidth]{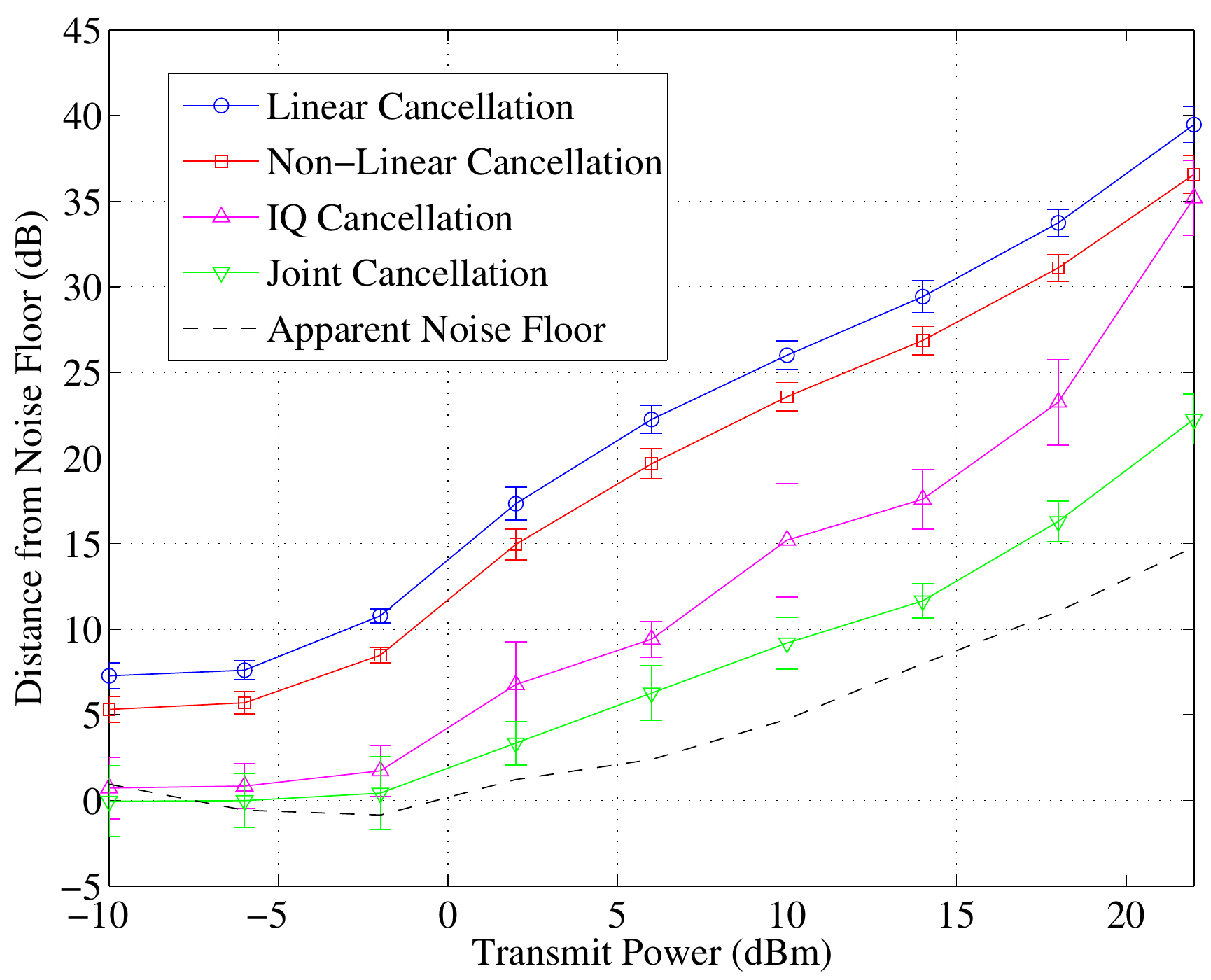}
	\caption{Distance from thermal noise floor as a function of the transmit power ($40$~dB analog cancellation).}\label{fig:NFdistance}
\end{figure}

In Fig.~\ref{fig:NFdistance} we present the mean distance of the residual self-interference signal from the measured noise floor for several digital cancellation methods with $40$~dB of passive suppression. Since the amount of cancellation is a random variable, we also include error bars at one standard deviation from the mean.  We also present the apparent noise floor at each transmit power, which stems mainly from the limited dynamic range of the ADC of the receiver. The achievable cancellation of any cancellation method is limited by this apparent noise floor and the only way to overcome it is to increase the amount of cancellation in the analog domain. We observe that simple linear cancellation alone is insufficient even at low transmit powers, as the residual signal lies at least $7$~dB above the noise floor. Non-linear cancellation~\cite{Bharadia2013} only reduces the residual self-interference by an additional $2$~dB over the entire examined range of transmit powers. The IQ imbalance cancellation method proposed in~\cite{Korpi2014} improves the obtained cancellation drastically. However, our joint cancellation method consistently outperforms all previously proposed methods as it considers both the baseband non-linearities and the IQ imbalance.

\begin{figure}[t]
	\centering
	\includegraphics[width=0.44\textwidth]{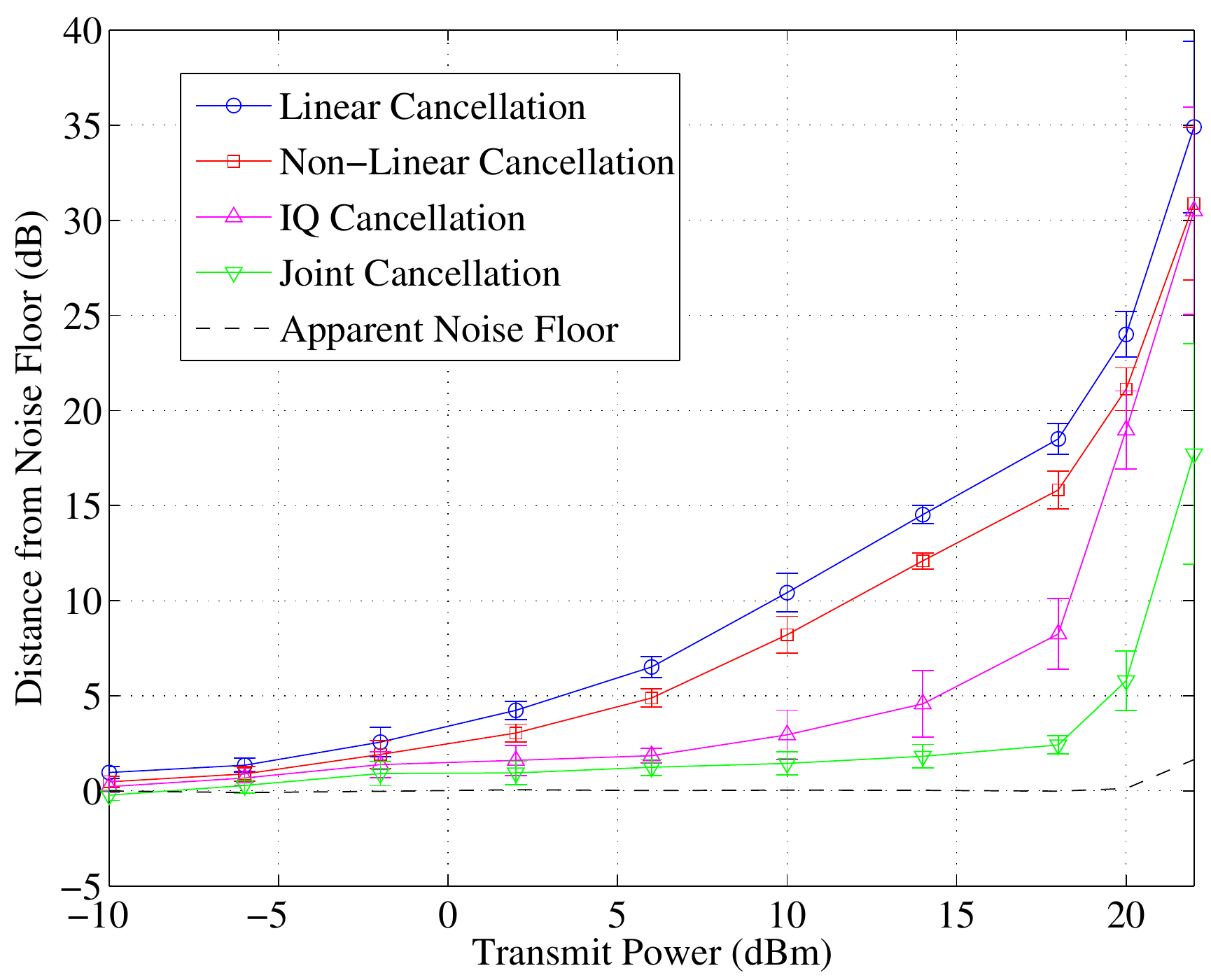}
	\caption{Distance from thermal noise floor as a function of the transmit power ($55$~dB analog cancellation).}\label{fig:NFdistance2}
\end{figure}

In Fig.~\ref{fig:NFdistance2} we present the mean distance of the residual self-interference signal from the measured noise floor for several digital cancellation methods with $55$~dB of passive suppression. In this case, the apparent noise floor remains equal to the measured thermal noise floor for all considered transmit powers. Moreover, we observe that at low transmit powers almost all cancellation methods perform equally well, because the required amount of suppression is relatively low and most of the non-idealities lie below the thermal noise floor. As the transmit power is increased, however, IQ imbalance cancellation and our joint cancellation perform significantly better than linear and non-linear cancellation. It is important to note that our joint cancellation method manages to keep the residual self-interference at less than $3$~dB above the noise floor up to a transmit power of $18$~dBm. However, all cancellation methods start failing at transmit powers above $18$~dBm.

\subsubsection{High RF Output Power}
In this section, we extend our non-idealities model to include RF non-linearities, providing some clues as to why the joint cancellation method starts failing at high transmit powers.

When the output power is high, the upconverted analog baseband signal $\ctpbx$ is unaltered with respect to the low RF power case. However, the transmitter amplifier introduces non-linearities in its output signal $\ctpbxamp$. In \eqref{eqn:txampnl}, we use the Taylor series expansion of \eqref{eqn:ampnl} to model the amplified analog RF signal. At the receiver, the RF mixer introduces IQ imbalance and the received analog baseband signal $\ctpbrx$ becomes:
\begin{align}
	\ctpbrx	& = \text{LPF}\left\{\left(\rxmixergain \ast \ctpbxamp + \rxmixergainconj \ast \ctpbxamp ^*\right) e^{-j2\pi f_ct}\right\} \\
					& = \left(\frac{\rxmixergain+\rxmixergainconj}{2}\right) \ast \sum _{\substack{n=1,\\n~\text{odd}}}^{n_{\max}}\frac{\ampnl{n}}{2^{n-1}} {n \choose \frac{n-1}{2}} \nonumber\\ 
					& \times \left(\txmixergain \ast \ctbbx + \txmixergainconj \ast \ctbbx ^*\right)^{\frac{n+1}{2}} \left(\txmixergain^* \ast \ctbbx^* + \txmixergainconj ^* \ast \ctbbx \right)^\frac{n-1}{2} \label{eqn:nonidealitieshighpower}
\end{align}
Thus, at high RF output power the received baseband signal $\dtbbrx$ contains products between $\ctbbx$ and $\ctbbx^*$, or equivalently, $\Re\{\ctbbx\}^m$ and $\Im\{\ctbbx\}^m,~m = 1,\hdots,m_{\max}$, which are not contained in the model of \eqref{eqn:nonidealitieslowpower:2} and are thus not compensated for properly.

In principle, the model in \eqref{eqn:nonidealitieshighpower} can be used for LS-based non-linear digital cancellation similarly to the one in \eqref{eqn:nonidealitieslowpower:2}. However, the number of terms in $\eqref{eqn:nonidealitieshighpower}$ and, consequently, the number of channels that need to be estimated grows as $(2m_{\max})^{n_{\max}}$, which is prohibitively large even for small values of $m_{\max}$ and $n_{\max}$. For example, for $m_{\max} = n_{\max} = 3$ we have $(2m_{\max})^{n_{\max}} = 216$. The joint cancellation of DAC non-linearities, IQ imbalance, and amplifier non-linearities remains a challenging open problem.

\section{Conclusion}
Self-interference cancellation in full-duplex systems is limited by the presence of hardware impairments in the transmitter and receiver circuits. In this paper, we provided a measurement-based study of the transceiver impairments that play a significant role in full-duplex wireless systems.  Our measurements confirmed the existence of several impairments only previously considered analytically, such as phase-noise and IQ imbalance, but they also demonstrated the existence of significant DAC induced baseband non-linearities. Measurements from our testbed show that our digital cancellation method that jointly takes into account DAC non-linearities and IQ imbalance achieves up to $13$~dB more self-interference cancellation than existing digital cancellation methods.

\section*{Appendix - DAC Non-Linearities}
Let $\omega \triangleq 2 \pi f$ denote the transmitted tone frequency so that for the ideal complex analog baseband signal we have:
\begin{align}
	\ctbbx	& = e^{j\omega t} = \cos (\omega t) + j \sin (\omega t).
\end{align}
Assume that DACs are perfectly matched so that $\dacnl{1}{m} = \dacnl{2}{m} = \alpha _m,~m \in \{1,\hdots,m_{\max}\}$. Then, the non-ideal complex analog baseband signal with DAC induced non-linearities is:
\begin{align}
	\ctbbx	& = \sum _{m=1}^{m _{\max}} \alpha _m\Re\{\ctbbx\}^m + j\sum _{m=1}^{m _{\max}} \alpha _m\Im\{\ctbbx\}^m\\
					& = \sum _{m=1}^{m _{\max}} \alpha _m\cos ^m (\omega t) + j\sum _{m=1}^{m _{\max}} \alpha _m\sin ^m (\omega t).
\end{align}
Let $\omega _c \triangleq 2 \pi f_c$ denote the carrier frequency. Assuming an ideal RF mixer, the analog RF signal is given by:
\begin{align}
	\ctpbx 	& = \cos(\omega _c t) \sum _{m=1}^{m _{\max}} \alpha _m \cos^m(\omega t) \nonumber \\
					& + \sin(\omega _c t) \sum _{m=1}^{m _{\max}} \alpha _m\sin^m(\omega t). \label{eqn:rf}
\end{align}
We define $\omega _{m,k} \triangleq (m-2k)\omega$ and $s_{m,k} \triangleq (-1)^{(\frac{m-1}{2}-k)}$. There are two cases for $m$. When $m$ is \emph{odd}, we have:
\begin{align}
	\cos^m(\omega t) & = \frac{2}{2^m} \sum_{k=0}^{\frac{m-1}{2}} \binom{m}{k} \cos{(\omega _{m,k} t)}, \label{eqn:cosm} \\
	\sin^m(\omega t) &= \frac{2}{2^m} \sum_{k=0}^{\frac{m-1}{2}} s_{m,k} \binom{m}{k} \sin{(\omega _{m,k} t)}.  \label{eqn:sinm}
\end{align}
Note that, since $\frac{m-1}{2}$ is always even, we have:
\begin{align}
	s_{m,k}	& = \left\{ \begin{tabular}{ll} $+1$, & $k$ even, \\ $-1$, & $k$ odd. \end{tabular}  \right.
\end{align} 
By replacing \eqref{eqn:cosm} and \eqref{eqn:sinm} in \eqref{eqn:rf}, we get:
\begin{align}
	\ctpbx 	& = \cos(\omega _c t) \sum _{m=1}^{m _{\max}}\frac{\alpha _m}{2^{m-1}} \sum_{k=0}^{\frac{m-1}{2}} \binom{m}{k} \cos{(\omega _{m,k}t)} \nonumber \\
					& + \sin(\omega _c t) \sum _{m=1}^{m _{\max}}\frac{\alpha _m}{2^{m-1}} \sum_{k=0}^{\frac{m-1}{2}} s_{m,k} \binom{m}{k} \sin{(\omega _{m,k}t)} \\
				 	& = \sum _{m=1}^{m _{\max}}\frac{\alpha _m}{2^{m}} \sum_{k=0}^{\frac{m-1}{2}} \binom{m}{k} (1+s_{m,k})\cos{(\omega_c - \omega _{m,k}t)} \nonumber \\
					& +  (1-s_{m,k})\cos{(\omega_ct + \omega _{m,k}t)}.
\end{align}
Thus, when $s_{k,m} = +1$ all $\cos{(\omega_ct + \omega _{m,k}t)}$ terms disappear. On the other hand, when $s_{k,m} = -1$, all $\cos{(\omega_ct - \omega _{m,k}t)}$ terms disappear. Thus, with ideal downconversion, the harmonics resulting from odd values of $m$ appear at frequencies $m(-1)^{\frac{m-1}{2}}\omega$ in the analog baseband. 

Let $c_{m} \triangleq \frac{1}{2^m} \binom{m}{\frac{m}{2}}$. When $m$ is \emph{even}, we have:
\begin{align}
	\cos^m(\omega t) 	& = c_{m} + \frac{2}{2^m} \sum_{k=0}^{\frac{m}{2}-1} \binom{m}{k} \cos{(\omega _{m,k}t)}, \label{eqn:cosmeven} \\
	\sin^m(\omega t)		& = c_{m} + \frac{2}{2^m} \sum_{k=0}^{\frac{m}{2}-1} s_{m+1,k}\binom{m}{k} \cos{(\omega _{m,k}t)}. \label{eqn:sinmeven}
\end{align}
By replacing \eqref{eqn:cosmeven} and \eqref{eqn:sinmeven} in \eqref{eqn:rf} it can be shown similarly that no terms cancel out. Thus, with ideal downconversion, the harmonics resulting from even values of $m$ appear at both $-m \omega$ and $+m \omega$ in the analog baseband signal.


\begin{backmatter}




\bibliographystyle{bmc-mathphys} 
\bibliography{bmc_article}      

\end{backmatter}
\end{document}